\documentclass[sigconf,nonacm]{acmart}

\usepackage{lmodern}
\microtypesetup{expansion=false}
\usepackage{tikz}
\usetikzlibrary{positioning,arrows.meta,fit,backgrounds,calc}

\usepackage{seqsplit}

\usepackage{booktabs}
\usepackage{xcolor}
\usepackage{listings}

\newcommand{\tool}{\textsc{uringscope}}

\setcopyright{none}
\renewcommand\footnotetextcopyrightpermission[1]{}
\AtBeginDocument{\setlength{\parskip}{0pt}}
\begin{document}

\title{\tool: Portable, Low-Overhead Observability for io\_uring}

\author{Rajarshi Chowdhury}
\orcid{0009-0007-7032-2450}
\affiliation{%
  \institution{\href{https://orcid.org/0009-0007-7032-2450}{orcid.org/0009-0007-7032-2450}}
  \country{}%
}

\begin{abstract}
io\_uring moves I/O submission and completion into shared-memory rings.
This makes it fast, and it also makes it invisible. \texttt{strace} sees
only the ring setup, and the kernel tracepoints that expose the request flow
are not stable ABI, so the few tools built on them work only on narrow
kernel ranges. We present \tool, a single-binary, language-agnostic
observability tool for io\_uring built on CO-RE (Compile Once, Run
Everywhere) eBPF. \tool{} makes four contributions. The first is a precise model of the
request lifecycle and a method to reconstruct per-request flows from kernel
events. The second is a technique for attaching portably to an unstable
tracepoint surface, using BTF-probed program variants, CO-RE field flavors,
and position-independent reads. The third is an evaluation of the tradeoff
between overhead and fidelity: on device-bound NVMe workloads \tool's
aggregate mode costs 0.7 to 9.9\% of throughput, which is cheaper than every
full-fidelity alternative we measured. The fourth is a lightweight
correctness mode that reuses the same reconstruction to detect
submission-boundary hazards, together with a built-in \emph{doctor} that
turns the measurements into named pathologies with evidence, for operators
who are debugging a tail-latency incident rather than browsing histograms.\footnote{Source code and the evaluation artifacts that
reproduce this paper's figures and tables are archived at
\url{https://doi.org/10.5281/zenodo.20672341}.}
\end{abstract}

\maketitle

\section{Introduction}

For two decades, the answer to ``what is my process doing to the kernel?''
has started with \texttt{strace}. io\_uring breaks that contract by design.
Submission queue entries (SQEs) and completion queue entries (CQEs) travel
through memory-mapped rings, and a busy application may issue millions of
I/O requests through a handful of \texttt{io\_uring\_enter} calls, or none
at all when it uses \texttt{SQPOLL}. Tracing the syscall boundary of such an
application yields a stream of opaque \texttt{io\_uring\_enter} calls and
none of the requests inside them. This limitation was raised on the strace
mailing list in 2020 and remains unaddressed~\cite{strace-iouring-thread}.

The information is not hidden. It is merely unpackaged. The kernel exposes
roughly eighteen static tracepoints covering the io\_uring request
lifecycle~\cite{kernel-iouring-tracepoints}, plus ring state in
\texttt{/proc/<pid>/fdinfo}. Practitioners who need answers today are
pointed at raw \texttt{tracefs}, \texttt{perf}, and hand-written
\texttt{bpftrace} scripts~\cite{cloudflare-iowq}. The one packaged tool,
\texttt{uring-trace}~\cite{uring-trace}, demonstrated the demand but also
the obstacle. Its authors describe the difficulty of building a stable tool
on top of unstable interfaces, and it supports only kernels 6.1 through 6.7.
io\_uring's tracepoints are explicitly not kernel ABI. They have been
renamed, re-prototyped, and re-fielded repeatedly across the 5.15 to 6.x
span (\S\ref{sec:churn}).

This paper argues that the gap is closable, and that closing it surfaces two
problems of independent interest. The first is a semantic problem. Mapping
kernel events back to application-meaningful request flows requires a precise
model of the io\_uring request lifecycle: when requests complete inline, when
they park on the poll-retry path, when they silently fall back to the
\texttt{io-wq} asynchronous worker pool (a classic hidden source of tail
latency), and how multishot and linked requests break the
one-submit-one-completion assumption. To our knowledge this model exists only
as kernel source. We make it explicit (\S\ref{sec:lifecycle}) and build
\tool's reconstruction on it. The second is a mechanical problem: attaching
to a tracepoint surface that changes across kernel versions. CO-RE
relocations solve struct-layout drift but not renamed or re-prototyped
tracepoints. \S\ref{sec:churn} contributes a layered technique that combines
struct-centric reads, BTF-probed program variants, CO-RE flavors, and a
kernel-matrix CI that treats portability as a continuously tested claim. We
position this as a constructive counterpart to DepSurf's measurement of the
eBPF dependency-surface problem~\cite{depsurf}.

\tool{} packages both into a single static binary, invoked as
\texttt{uringscope ./myapp}. Its default mode aggregates per-opcode latency
histograms and pathology counters entirely in kernel maps, which is cheap
enough to leave on in production. A trace mode streams per-request lifecycle
records into a Perfetto-compatible timeline~\cite{perfetto}. A \emph{doctor}
layer converts the measurements into verdicts with evidence and suggested
fixes.

\noindent Contributions:
\begin{itemize}
\item A request-lifecycle model for io\_uring and a per-request flow
      reconstruction that handles multishot operations, linked SQEs, SQPOLL
      submission, and deferred task-work delivery (\S\ref{sec:lifecycle}).
\item A portable-attachment technique for unstable tracepoint surfaces,
      with a churn study of io\_uring's tracepoints across 5.15--6.x
      (\S\ref{sec:churn}).
\item \tool{} itself, open source, plus an overhead-vs-fidelity evaluation
      against \texttt{perf}, \texttt{bpftrace}, and \texttt{strace} at up
      to saturation IOPS (\S\ref{sec:eval}).
\item A lightweight \emph{correctness} mode that reuses the lifecycle
      reconstruction to detect submission-boundary hazards. These are
      overlapping in-flight operations on a buffer, registered-buffer
      lifetime violations, dropped requests that are never reaped, and the
      unmap variant of buffer use-after-free. We give an explicit account of
      which hazards are and are not detectable from the kernel side
      (\S\ref{sec:hazards}).
\end{itemize}

\section{Background and Motivation}
\label{sec:background}

\subsection{io\_uring in one column}
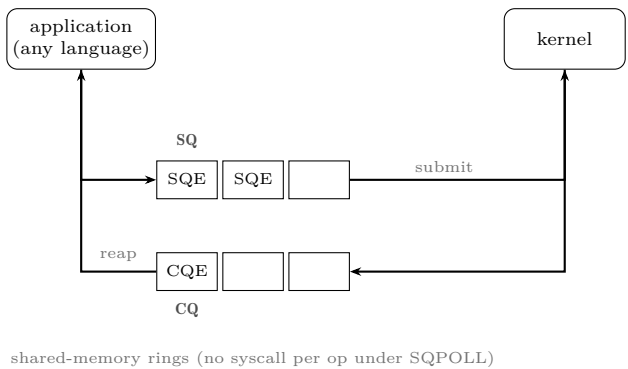
\begin{figure}
\centering
\footnotesize
\begin{tikzpicture}[
  font=\footnotesize,
  box/.style={draw, rounded corners, minimum width=16mm, minimum height=8mm, align=center, inner sep=2pt},
  ring/.style={draw, minimum width=8mm, minimum height=5mm, inner sep=1pt, font=\scriptsize},
  lbl/.style={font=\scriptsize, text=black!55},
  rowlbl/.style={font=\scriptsize\bfseries, text=black!70},
  ar/.style={-{Stealth[length=1.5mm]}, thick}
]
\node[box] (app) {application\\(any language)};
\node[box, right=46mm of app] (kern) {kernel};

\node[ring, below=12mm of app, xshift=14mm] (sq1) {SQE};
\node[ring, right=0.6mm of sq1] (sq2) {SQE};
\node[ring, right=0.6mm of sq2] (sq3) {\phantom{SQE}};
\node[ring, below=7mm of sq1] (cq1) {CQE};
\node[ring, right=0.6mm of cq1] (cq2) {\phantom{CQE}};
\node[ring, right=0.6mm of cq2] (cq3) {\phantom{CQE}};

\node[rowlbl, above=0.6mm of sq1] {SQ};
\node[rowlbl, below=0.6mm of cq1] {CQ};

\draw[ar] (app.south) |- (sq1.west);
\draw[ar] (sq3.east) -| node[lbl, pos=0.22, above] {submit} (kern.south);
\draw[ar] (kern.south) |- (cq3.east);
\draw[ar] (cq1.west) -| (app.south);
\node[lbl, above=0.4mm] at ([xshift=-5mm]cq1.west) {reap};

\node[lbl, below=7mm of cq2] {shared-memory rings (no syscall per op under SQPOLL)};
\end{tikzpicture}
\caption{io\_uring's shared rings. The application writes SQEs to the
submission queue and reads CQEs from the completion queue. The request flow
lives in this shared memory.}
\label{fig:rings}
\end{figure}

io\_uring~\cite{axboe-iouring} replaces the synchronous syscall boundary
with two memory-mapped rings shared between the application and the kernel.
The application fills a submission queue (SQ) with submission queue entries
(SQEs), and the kernel fills a completion queue (CQ) with completion queue
entries (CQEs). One \texttt{io\_uring\_enter} syscall can submit a batch of
SQEs and reap many CQEs. With \texttt{SQPOLL}, a kernel thread polls the SQ
and the application issues no syscalls at all. The libc-level wrapper
\texttt{liburing} hides the ring bookkeeping (Figure~\ref{fig:rings}).
Four features matter for observation. \emph{Registered} files and buffers
are referenced by index rather than by fd or pointer. \emph{Multishot}
operations post many CQEs for one SQE. \emph{Linked} SQEs form ordered
chains. And \texttt{DEFER\_TASKRUN} (6.1 and later) batches completion
delivery into ring-local task work. Each of these changes how a request
flows from submission to reaping, which is what an observer must reconstruct
(\S\ref{sec:lifecycle}).

\subsection{Why existing tools fail}
\texttt{strace} decodes the arguments to \texttt{io\_uring\_setup} and
\texttt{io\_uring\_register}, but the request flow lives in the shared rings,
which it never inspects. The 2020 strace-devel
thread~\cite{strace-iouring-thread} that raised this produced no fix. Our
measurements (\S\ref{sec:eval}) show that its syscall-stop machinery adds
44.4\,$\mu$s per request and distorts p99.9 latency by $33\times$, so it
changes the behavior it is meant to observe.

\texttt{perf record -e 'io\_uring:*'} and hand-written \texttt{bpftrace}
scripts read the same tracepoints \tool{} does, and for counting they work.
But they have three problems. They emit raw events with no request-level
correlation, so a \texttt{queue\_async\_work} event is just a name, not
``this READ for this user\_data took the slow path.'' They defer all
semantics to offline processing. And they collapse at high IOPS because
every event crosses to userspace (\S\ref{sec:eval-overhead}), and break
across kernels when a prototype changes (\S\ref{sec:churn}). The kernel also
exposes cheap complementary state in \texttt{/proc/<pid>/fdinfo}. \tool{}
parses it for ring geometry and SQPOLL identification (the SQ and CQ masks
and the SQPOLL thread id); instantaneous queue depth instead comes from a
CO-RE read of the ring's completion-queue head and tail
(\S\ref{sec:churn-validated}), and io-wq worker counts from a BPF map fed by
\texttt{sched\_switch}. fdinfo is real and useful, but it is a snapshot, not
a request history.

The one packaged tool, \texttt{uring-trace}~\cite{uring-trace}, is closest
in spirit, since it also uses eBPF with Perfetto output, and it demonstrates
the demand. But it is per-event by design and supports only kernels 6.1
through 6.7, which is the exact brittleness \S\ref{sec:churn} addresses.

\subsection{Who needs this}
The audience is specific and worth stating honestly. io\_uring is blocked
by Docker's default seccomp profile~\cite{docker-seccomp} and was disabled
fleet-wide at Google for attack-surface reasons~\cite{google-iouring}. The
relevant deployments are therefore bare-metal and VM storage engines,
databases, and io\_uring-native runtimes such as fio, TigerBeetle-class
systems, and the \texttt{tokio-uring}, \texttt{glommio}, and \texttt{eio}
ecosystems. This is a small population with a high blast radius. These are
precisely the systems where a hidden async-worker punt becomes a
customer-visible tail-latency regression, and where the engineer debugging
it today has no tool between \texttt{strace}, which is useless here, and
reading kernel source.

\section{The io\_uring Request Lifecycle}
\label{sec:lifecycle}

A request that an application thinks of as ``one read'' takes one of several
kernel paths, and which path it took is the whole story for tail latency. We
make the model explicit, and \tool's reconstruction is built on it.

\begin{figure}[t]
\centering
\footnotesize
\begin{tikzpicture}[
  node distance=9mm and 12mm,
  state/.style={draw, rounded corners, align=center, inner sep=3pt, font=\footnotesize, minimum height=6mm},
  ghost/.style={draw, dashed, rounded corners, align=center, inner sep=3pt, font=\footnotesize\itshape, text=black!55},
  tp/.style={font=\scriptsize\ttfamily, text=black!55, inner sep=1pt},
  ev/.style={-{Stealth[length=1.4mm]}, shorten >=0.5pt, shorten <=0.5pt}
]
\node[ghost] (prep) {SQE in SQ ring (unsubmitted)};
\node[state, below=7mm of prep] (issued) {issued};
\node[state, below=16mm of issued] (poll) {poll-armed};
\node[state, left=of poll] (done1) {completed};
\node[state, right=of poll] (punt) {io-wq punt};
\node[state, below=16mm of poll] (cq) {completed to CQ};
\node[ghost, below=7mm of cq] (ready) {CQE ready};
\node[ghost, below=7mm of ready] (reaped) {reaped};

\draw[ev] (prep) -- node[tp,right] {submit\_req} (issued);
\draw[ev] (issued) -- node[tp,above,pos=.55,sloped] {inline} (done1.north);
\draw[ev] (issued) -- node[tp,left] {poll\_arm} (poll);
\draw[ev] (issued) -- node[tp,above,pos=.55,sloped] {queue\_async\_work} (punt.north);
\draw[ev] (done1.south) -- node[tp,below,pos=.55,sloped] {} (cq.north west);
\draw[ev] (poll.south) -- node[tp,right,pos=.4] {reissue} (cq.north);
\draw[ev] (punt.south) -- (cq.north east);
\draw[ev] (cq) -- node[tp,right] {complete} (ready);
\draw[ev] (ready) -- node[tp,right] {reap (uprobe)} (reaped);
\node[tp, anchor=west] at ([xshift=3mm]cq.east) {$\to$ cqe\_overflow};
\node[tp, anchor=east] at ([xshift=-3mm]ready.west) {task\_work\_run};
\end{tikzpicture}
\caption{io\_uring request lifecycle. Solid nodes are kernel states with the
tracepoint that makes each transition observable. Dashed nodes are
userspace-only segments that fire no tracepoint, reachable only via
\texttt{liburing} uprobes. The three issue paths, inline, poll-armed, and
io-wq punt, are the fork that determines tail latency.}
\label{fig:lifecycle}
\end{figure}
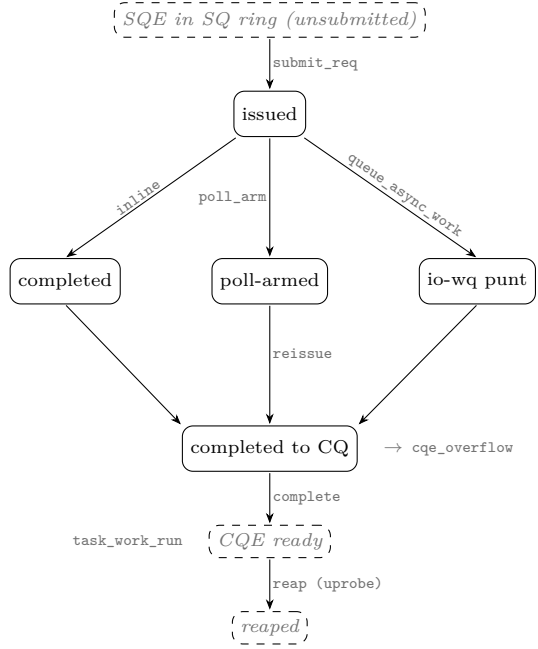

Figure~\ref{fig:lifecycle} is the model. After submission, the kernel issues
an SQE on one of three paths. On the inline path it completes immediately,
which is the fast path. On the poll-armed path it is pollable but not ready,
so it is parked and reissued. On the punt path it must block and cannot poll,
as with buffered I/O that misses the cache or an \texttt{fsync}, so it goes
to the io-wq worker pool. The punt is the classic hidden tail-latency source,
invisible to an application that sees only a slower completion. Completion
then posts a CQE, takes an overflow slow path if the CQ is full, and is
delivered via task work, batched under \texttt{DEFER\_TASKRUN}. Each
transition is exposed by exactly one tracepoint, named on the corresponding
edge.

Reconstructing per-request flow requires a correlation key. On modern
kernels the \texttt{io\_kiocb} pointer is stable from submit to completion
and is unique. Pre-5.19 legacy tracepoints force the weaker
\texttt{(ctx, user\_data)} pair, and \texttt{user\_data} is
application-chosen and freely reused, so it cannot be a sole key. Three
semantics complicate the simple submit-to-complete mapping. \emph{Multishot}
operations post many CQEs per SQE, flagged \texttt{IORING\_CQE\_F\_MORE}, so
\tool{} re-arms its in-flight record rather than retiring it, and latency
means the inter-completion gap. \emph{Linked} SQEs do not start until their
predecessor completes, so their submit-to-complete latency legitimately
includes queue-behind-link time. \emph{SQPOLL} performs submission on the
\texttt{iou-sqp} kernel thread, not in the application's syscall context.
This is why \tool{} filters by ring ownership, captured at
\texttt{io\_uring\_create}, rather than by the pid observed at the
tracepoint.

Two segments are fundamentally invisible to kernel tracepoints. One is the
gap between an application preparing an SQE and submitting it. The other is
the gap between a CQE becoming ready and the application reaping it. This
second gap, completion-reaping lag, is a real source of perceived latency.
\tool{} detects it from kernel-side completion-queue state, by spotting CQEs
that sit ready while unconsumed. Measuring the precise per-reap delay needs
optional \texttt{liburing} uprobes, which we discuss in
\S\ref{sec:limitations}.

\begin{table}
\centering
\small
\begin{tabular}{@{}lll@{}}
\toprule
signal & granularity & rate @ 1M IOPS \\
\midrule
\texttt{submit\_req}        & per request & 1M/s \\
\texttt{complete}           & per request & 1M/s \\
\texttt{queue\_async\_work} & per punt    & 0--1M/s \\
\texttt{poll\_arm}          & per not-ready & network-only \\
\texttt{task\_work\_run}    & per batch   & 10--100k/s \\
\texttt{enter} syscall      & per batch   & batching-dependent \\
\texttt{sched\_switch}      & system-wide & filtered in-probe \\
\bottomrule
\end{tabular}
\caption{Event rates per lifecycle signal. The two unavoidable per-request
tracepoints alone produce 2M events/s at 1M IOPS. Shipping that to userspace
is itself a workload, which is why \tool's default mode aggregates in the
kernel (\S\ref{sec:design}).}
\label{tab:rates}
\end{table}

\section{\tool{} Design}
\label{sec:design}

\subsection{Two modes, one dichotomy}
\begin{figure}
\centering
\resizebox{\columnwidth}{!}{%
\begin{tikzpicture}[
  font=\footnotesize,
  k/.style={draw, rounded corners, align=center, inner sep=2.5pt, minimum height=5.5mm},
  ob/.style={draw, rounded corners, align=center, inner sep=2.5pt, minimum height=5.5mm, fill=black!4},
  lbl/.style={font=\scriptsize, text=black!60},
  ar/.style={-{Stealth[length=1.4mm]}}
]
\node[k] (tp) {kernel tracepoints\\(per request)};
\node[k, above right=3mm and 14mm of tp] (maps) {BPF maps\\(histograms, counters)};
\node[k, below right=3mm and 14mm of tp] (rb) {ring buffer\\(per-event records)};
\node[ob, right=14mm of maps] (agg) {report + \emph{doctor},\\Prometheus};
\node[ob, right=14mm of rb] (perf) {Perfetto\\timeline};
\draw[ar] (tp.east) to[out=30,in=180] node[lbl, above, sloped, pos=0.5, yshift=0.6mm] {aggregate} (maps.west);
\draw[ar] (tp.east) to[out=-30,in=180] node[lbl, below, sloped, pos=0.5, yshift=-0.6mm] {trace} (rb.west);
\draw[ar] (maps) -- node[lbl,above] {at exit} (agg);
\draw[ar] (rb) -- node[lbl,above] {streamed} (perf);
\node[lbl, below=1.5mm of rb, xshift=-6mm] {bounded fidelity (drops counted)};
\node[lbl, above=1.5mm of maps, xshift=-4mm] {nothing per-event crosses};
\end{tikzpicture}}
\caption{Two modes from one tracepoint set. Aggregate mode keeps per-event
data in kernel maps, for production use. Trace mode streams records to a
timeline, for diagnosis. The split between in-kernel aggregation and
bounded-fidelity streaming is what the evaluation measures.}
\label{fig:modes}
\end{figure}
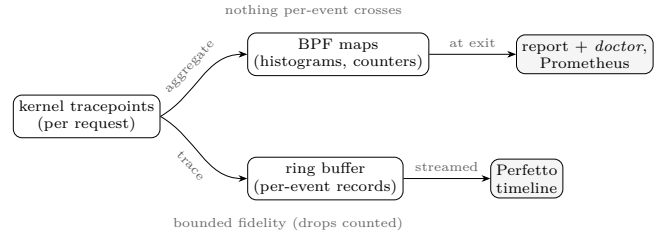

\tool{} offers two modes that trade fidelity for cost (Figure~\ref{fig:modes}),
and the choice between them is the backbone of the evaluation (\S\ref{sec:eval}).
\emph{Aggregate} mode folds everything into BPF maps in the kernel. These
hold per-opcode log2 latency histograms, the pathology counters for
Table~\ref{tab:rates}'s signals, SQPOLL off-CPU time, and io-wq fan-out.
Userspace reads the maps once at exit, or periodically for Prometheus export.
Nothing per-event crosses the boundary, so this is the always-on,
production-candidate mode. \emph{Trace} mode instead streams one per-request
record over a ring buffer to a Perfetto timeline. Its fidelity is bounded by
ring-buffer capacity, and \tool{} reports any drops as a first-class counter
rather than losing events silently. Reconstruction fidelity was 100\% in
both modes across our grid (\S\ref{sec:eval}).

\subsection{Correlation and state}
The hot structure is an in-flight hash keyed by the \texttt{io\_kiocb}
pointer (\S\ref{sec:lifecycle}). It stores a 48-byte record per outstanding
request: the submit and punt timestamps, opcode, \texttt{user\_data}, flags,
the owning \texttt{tgid}, and the target descriptor (address, length, buffer
index, and kind) consumed by the \texttt{--check} hazard tier
(\S\ref{sec:hazards}). Entries are deleted on completion. Some completions find no matching entry, because
the request was submitted before attach or evicted under pressure, and these
are counted as untracked. That count is a first-class fidelity metric the
report surfaces rather than hides. A separate ring-ownership map, populated
at \texttt{io\_uring\_create} and back-filled from \texttt{/proc/<pid>/fd}
when attaching to an already-running process, lets \tool{} filter to a target
ring correctly even under SQPOLL, where submission happens on a kernel
thread. SQPOLL and io-wq worker accounting reads \texttt{sched\_switch} and
filters on the \texttt{iou-sqp-} and \texttt{iou-wrk-} comm prefixes inside
the probe, so that only the relevant context switches reach a map.

\subsection{The doctor}
On top of the aggregates sits a \emph{doctor}, a set of conservative
predicates that each name a known pathology, show the supporting evidence,
and suggest a fix. The shipped rules cover the per-opcode punt ratio
(buffered-I/O storms), submission batching (SQEs per
\texttt{io\_uring\_enter}), SQPOLL off-CPU fraction, unbounded io-wq fan-out,
CQ overflow, short writes, deferred task-work delivery
(\texttt{DEFER-TW}), per-opcode error rate, and dropped requests that
are never reaped, plus tool-fidelity self-reports. The governing principle
is that a checker which cries wolf is ignored, so thresholds favor precision
over recall. The evaluation validates this stance with explicit
false-positive guards (\S\ref{sec:eval-detect}). The report, findings
included, is also emitted as machine-readable JSON under a versioned schema,
and a \texttt{--fail-on} severity threshold maps findings to deterministic
exit codes, so the doctor can gate CI runs and agent-driven workflows. The doctor is the feature that turns ``writes are
slow'' into ``\texttt{fsync} is punting 100\% to the bounded worker pool and
owns your p99'' (\S\ref{sec:eval}, case study).

\section{Surviving the Unstable Dependency Surface}
\label{sec:churn}

Kernel tracepoints are explicitly not ABI, and io\_uring's have exercised
that freedom repeatedly. Table~\ref{tab:churn} catalogs the changes \tool{}
handles across 5.15 to 6.17. DepSurf~\cite{depsurf} formalizes this
instability as a ``dependency surface,'' and this section is the
constructive counterpart, a tool engineered to survive it. CO-RE relocates
struct field offsets automatically, but it does nothing for renamed or
re-prototyped tracepoints, which need the layered technique below.

\begin{table}
\centering
\small
\footnotesize
\begin{tabular}{@{}p{3.05cm}cp{3.0cm}@{}}
\toprule
change & kernel & \tool's response \\
\midrule
\texttt{submit\_sqe} $\to$ \texttt{submit\_req} rename & 5.18--6.0 & two program variants, BTF-probed \\
\texttt{io\_kiocb.\allowbreak user\_data} into embedded \texttt{io\_cqe} & 5.19 & CO-RE flavor + \texttt{field\_exists} \\
\texttt{cqe\_overflow} added & 5.19 & probe-by-name, absent {=} off \\
\texttt{short\_write} added & 6.0 & probe-by-name \\
\texttt{local\_work\_run} added & 6.1 & probe-by-name \\
\texttt{complete} args $\to$ \texttt{io\_uring\_cqe~*} & 6.17 & third variant, read via cqe \\
\bottomrule
\end{tabular}
\caption{io\_uring tracepoint churn handled by \tool{}, with the mechanism
for each. The 6.17 \texttt{complete} collapse was encountered live on the
measurement host (\S\ref{sec:eval}). Format dumps from each kernel are the
table's primary sources.}
\label{tab:churn}
\end{table}

\noindent Then the technique, layered by what changed:
\begin{itemize}
\item \textbf{Fields moved} $\Rightarrow$ CO-RE flavors +
      \texttt{bpf\_core\_field\_exists()} (one compiled program serves both
      layouts).
\item \textbf{Prototypes reshuffled} $\Rightarrow$ struct-centric reads:
      trust the \texttt{io\_kiocb} pointer, distrust argument positions.
\item \textbf{Tracepoints renamed/added/removed} $\Rightarrow$ multiple
      compiled variants. Userspace probes kernel BTF for
      \texttt{btf\_trace\_*} symbols and flips autoload per variant, so a
      missing tracepoint degrades one feature instead of failing load.
\item \textbf{The future} $\Rightarrow$ a nightly kernel-matrix CI (5.15,
      6.1, 6.6, 6.8, 6.12, and mainline, the last currently supplying 6.17
      coverage) that boots each kernel, runs a known
      workload, asserts the expected support tier, and archives tracepoint
      formats, making portability a regression-tested property.
\end{itemize}

Concretely, the portability machinery is a small fraction of the tool. The
BPF object defines 18 kernel-attached programs (plus two \texttt{liburing}
uprobes for the userspace boundary, 20 in all) across 1{,}340 lines, and the
variant-selection logic in userspace is 412 lines. Four of the programs are
variants of a single tracepoint, \texttt{io\_uring\_complete}: the modern
5-argument form, the pre-6.0 legacy form, the 6.17 cqe-collapsed form, and a
count-only fallback. Two more are submit-path variants, \texttt{submit\_req}
versus the legacy \texttt{submit\_sqe}, and the remaining programs attach
once. At load time, nine autoload decisions driven by BTF probing select the
variants appropriate to the running kernel and disable the rest, and two
further hazard hooks are enabled by tracefs probing, for eleven rows in the
startup support summary. The cost of
spanning 5.15 to 6.17 is thus a handful of alternative programs and a
probe table, not a rewrite per kernel.

\subsection{Empirical validation of the technique}
\label{sec:churn-validated}
The mechanism is not hypothetical. On a stock 6.6 kernel, libbpf load-time
logs confirm each layer firing as designed. The 5.19 field move, in which
\texttt{io\_kiocb}'s \texttt{user\_data} migrated into an embedded
\texttt{io\_cqe}, is handled live. The modern relocation resolves
\texttt{io\_kiocb.cqe.user\_data} to its kernel offset, while the
\texttt{io\_kiocb\_\_\_pre519} flavor finds no match and its load is poisoned
by \texttt{bpf\_core\_field\_exists()}, which is exactly the intended
degrade-not-fail behavior. The struct-centric instantaneous-depth read
relocates \texttt{io\_ring\_ctx.rings} to \texttt{io\_rings.cq.\{head,tail\}}
without touching tracepoint arguments. BTF probing then disables the legacy
\texttt{submit\_sqe} and \texttt{complete} variants on this kernel and
enables the modern \texttt{submit\_req} path. The startup support summary on
the measurement kernel makes the selection legible. It reports each feature
as active with the variant chosen, for example \texttt{completion: active,
v6.17 cqe-collapsed (3-arg)} and \texttt{submission: active, modern
submit\_req}, and it disables the legacy variants. Table~\ref{tab:churn}
lists the per-change mechanism, and the full multi-kernel coverage matrix is
left to the CI described in \S\ref{sec:limitations}.

\section{Correctness: Hazards at the Submission Boundary}
\label{sec:hazards}

io\_uring inverts the ownership contract of synchronous I/O. A buffer handed
to the kernel at submit is not the application's again until the matching
completion. Four hazards live in that window. The first is use-after-free or
\texttt{munmap} of an in-flight buffer. The second is two in-flight
operations writing the same buffer range, which silently corrupts data and
never returns an error. The third is a registered-buffer lifetime violation,
where an index is unregistered or re-registered while it has live references.
The fourth is a dropped request that is submitted but never reaped, which
pins the buffer forever. The same lifecycle tracking that powers observation
makes a lightweight correctness checker possible, a ``valgrind for the
io\_uring submission boundary.''

The contribution here is as much epistemic as mechanical. It is a precise
statement of which hazards are detectable from which vantage point
(Table~\ref{tab:hazards}). Hazards 2, 3, and 4, together with the
\texttt{munmap} variant of hazard 1, are reachable from kernel tracepoints
alone, because \tool{} already tracks every in-flight request keyed by
\texttt{io\_kiocb}. The freelist or stack-reuse variant of hazard 1 fires no
syscall and is invisible without allocator instrumentation. We disclaim it
explicitly rather than overclaim.

\begin{table}
\centering
\small
\footnotesize
\begin{tabular}{@{}llc@{}}
\toprule
hazard & vantage & status \\
\midrule
dropped requests (4)        & kernel tracepoints & shipped \\
overlapping in-flight (2)   & kernel tracepoints & shipped \\
registered-buf lifetime (3) & + register tp     & shipped \\
UAF, \texttt{munmap} variant (1) & + \texttt{munmap} tp & shipped \\
UAF, free/stack variant (1) & needs allocator hooks & disclaimed \\
reaping lag (CQEs ready)    & kernel tracepoints & shipped \\
\bottomrule
\end{tabular}
\caption{Submission-boundary hazards by detectability. The honest line
between what kernel-side tracing can and cannot catch is itself a result.}
\label{tab:hazards}
\end{table}

\tool's \texttt{--check} mode implements these as a debugging tier, an
``ASan for the io\_uring submission boundary'' rather than the always-on
path. It maintains per-ring in-flight target ranges, using address and
length for plain operations and \texttt{buf\_index} and offset for fixed
operations, captured at submit by CO-RE-reading the \texttt{io\_kiocb}. From
these it runs four checks. The first is an overlap test on submit (hazard 2).
The second validates \texttt{buf\_index} reference counts against the
register tracepoint (hazard 3). The third tests \texttt{munmap} ranges
against in-flight targets (hazard 1, unmap variant). The fourth is an aged
in-flight scan for dropped requests (hazard 4, always available). All four
are validated by injection in \S\ref{sec:eval-detect}.

The BPF verifier shapes this design. The overlap test cannot iterate the
in-flight hash map, because the verifier rejects unbounded loops. \tool{}
therefore keeps a fixed-size per-ring scan window of the $K$ most recent
in-flight target descriptors and tests new submissions against it with a
\texttt{\#pragma unroll}ed comparison. We set $K$ to 64. Hazard-2 detection
is thus a bounded approximation. An overlap against an in-flight request older than
the last $K$ on the same ring is not caught by this rule, although the
dropped-request rule (hazard 4) still flags the very old stragglers. We
state this limit explicitly rather than imply exhaustiveness. For a
debugging-mode checker, last-$K$ coverage with a documented bound is the
honest and verifier-feasible point. A hazard is surfaced via an at-exit
counter-plus-samples map, the same transport as the dropped-request rule. It
carries both colliding \texttt{user\_data} tokens, both opcodes, and the
overlapping range, so a developer can grep their own code for the two
requests involved. In our implementation the in-kernel bounded scan held at
$K=64$ and passed the verifier on Linux 6.17. All three kernel-side hazard
rules, namely overlapping in-flight, registered-buffer lifetime, and the
\texttt{munmap} variant of use-after-free, detect their injected hazards with
no false positives (\S\ref{sec:eval-detect}). On the overlap injection the
doctor reports, verbatim, \emph{``READ\_FIXED(user\_data=0x41) and
READ\_FIXED(user\_data=0x42) overlap in registered buffer \#0 at
[\ldots,+4096)''}, alongside \texttt{[HAZARD-BUFREG]} flagging the subsequent
unregister of an index with two live references. The two detectors fire
together on one workload.

Detection effectiveness is scored by the injection harness of
\S\ref{sec:eval-detect}. This positions \tool{} distinctly from
RingGuard~\cite{ringguard}, which applies io\_uring+eBPF to security
\emph{policy enforcement}. Correctness checking of the buffer-ownership
contract is a separate and, to our knowledge, unoccupied point.

\section{Evaluation}
\label{sec:eval}

\subsection{Implementation status}
\label{sec:status}
Before the quantitative results, we state what is built and validated, so
that claims and measurements are not conflated. All eleven
injection-validated doctor rules detect their injected pathology on the
measurement kernel (Linux 6.17.0-1017), and the false-positive guards hold,
as detailed in \S\ref{sec:eval-detect}. The shipped rule set is larger than
the injection-validated eleven (\S\ref{sec:design}). Detection effectiveness is a correctness property,
validated independently of the performance campaign.
The performance results below come from a 36-cell grid of 6 workloads by 6
observers, with 5 runs per cell and medians reported, collected on the host
described in \S\ref{sec:setup}.

\subsection{Experimental setup}
\label{sec:setup}
The protocol is implemented in \texttt{bench/}. It is a grid over workloads,
observers, and load levels, with at least 5 runs per cell, fio pinned to a
fixed core set, and cold-cache control via \texttt{drop\_caches}. The
observers are nothing (baseline), \tool-aggregate, \tool-trace, \texttt{perf
record -e 'io\_uring:*'}, a \texttt{bpftrace} counter one-liner, and
\texttt{strace -c}. The workloads are the O\_DIRECT fast path, a buffered
cold-cache punt storm, an fsync writer, SQPOLL, iodepth-1, and a network echo
server. The five storage workloads are fio-driven; the network echo cell is
instead driven by \texttt{nc} echo streams against the server.

We run on a single host: Intel Xeon Platinum 8375C (Ice Lake, 2.9\,GHz),
32 logical CPUs (16 cores $\times$ 2 SMT), 256\,GiB RAM, Ubuntu 24.04,
\texttt{CONFIG\_DEBUG\_INFO\_BTF=y}. The CPU-bound (\texttt{null\_blk})
workloads ran on Linux 6.17.0-1017, and the device-bound NVMe workloads on
6.17.0-1012, an earlier point release of the same series. The AWS kernel
advanced mid-campaign, which is itself a minor illustration of the moving
target \S\ref{sec:churn} addresses. The physical
storage device is a datacenter NVMe SSD whose ceiling we report from
measurement rather than a part number: 322k 4\,KiB random-read IOPS at
QD64, p99 completion latency 232\,$\mu$s, p99.99 255\,$\mu$s, at 99.5\%
device utilization. Two honesty notes. First, the upper-bound overhead
numbers are measured against an in-kernel \texttt{null\_blk} device: this
removes storage from the critical path and makes the workload CPU-bound,
so observer cost translates directly into throughput loss. This is a
deliberate worst case. The physical NVMe carries the workloads where real-device
semantics matter (cold-cache reads, fsync) and represents the
device-bound production regime. Second, the host is a virtualized
instance (thin hypervisor): C-states and turbo are not pinned by the
guest. We mitigate by anchoring the worst-case claim on \texttt{null\_blk}
(largely jitter-independent), reporting medians of 5 runs per cell, and
noting a bare-metal re-run as future work. Each cell is a 60\,s run after
10\,s warmup, fio pinned to 4 cores, observers unpinned on the remaining 28.

\subsection{Overhead}
\label{sec:eval-overhead}

\begin{figure*}
\centering
\includegraphics[width=\textwidth]{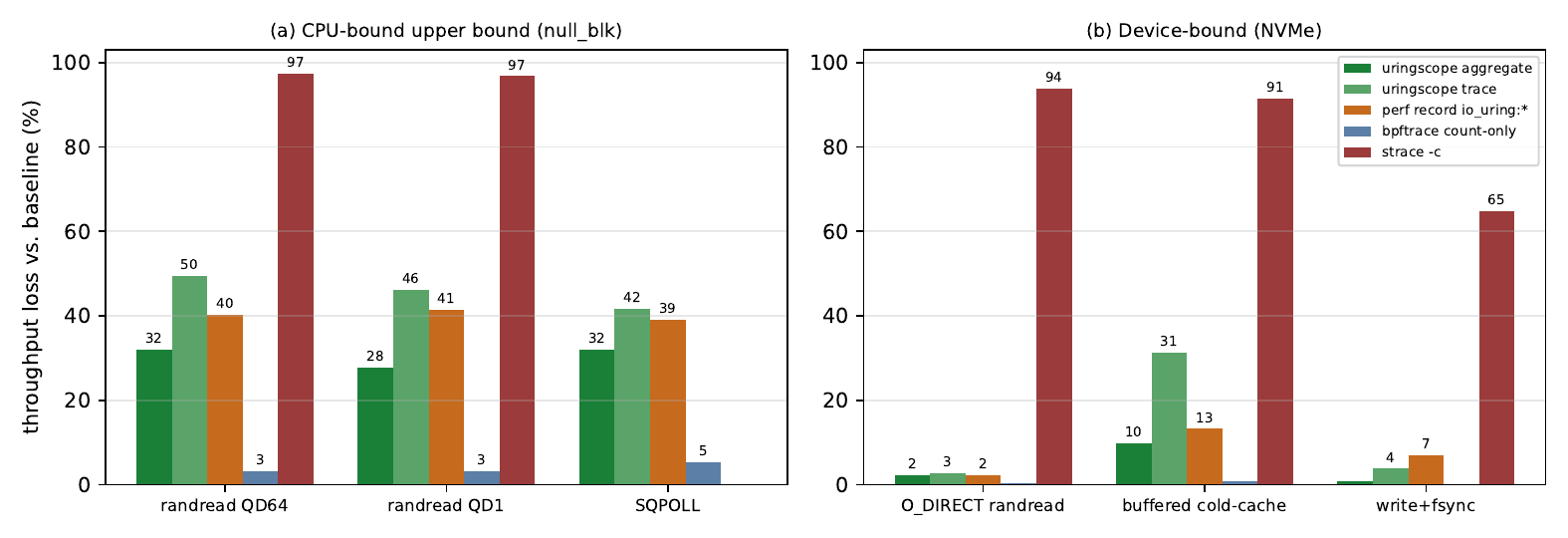}
\caption{Throughput loss per observer. (a)~CPU-bound \texttt{null\_blk}
workloads are the deliberate worst case, where every nanosecond of observer
cost subtracts from throughput. (b)~On the device-bound NVMe workloads that
represent production, \tool's aggregate mode costs 0.7 to 9.9\% and is the
cheapest full-fidelity observer on every workload. Medians of 5 runs.}
\label{fig:overhead}
\end{figure*}

\begin{figure}
\centering
\includegraphics[width=\columnwidth]{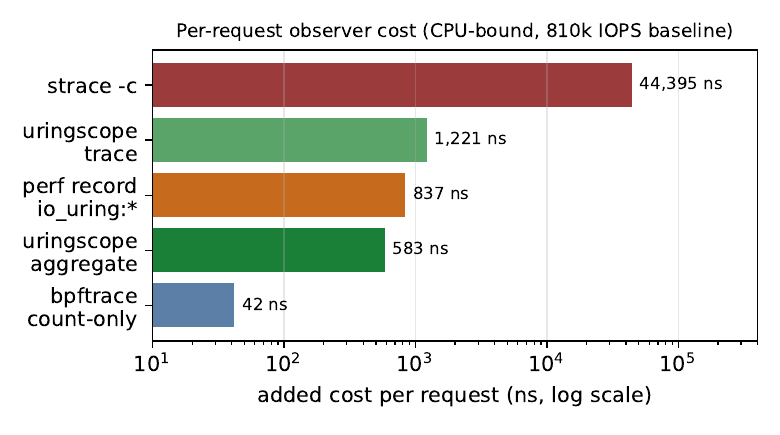}
\caption{Per-request observer cost, derived from inverse-throughput deltas
at the CPU-bound saturation point (810k IOPS baseline). This is the
load-independent way to state overhead. Multiply by your IOPS to predict
throughput cost when CPU-bound. When device-bound, idle headroom absorbs it.}
\label{fig:perreq}
\end{figure}

Figure~\ref{fig:overhead} shows throughput loss per observer, and
Figure~\ref{fig:perreq} restates the CPU-bound case as cost per request,
which is the durable number. There are three results.

First, device-bound workloads, which are the production case, are cheap to
observe. On the NVMe O\_DIRECT fast path (330k IOPS), \tool-aggregate costs
2.2\%, on write+fsync it costs 0.7\%, and on buffered cold-cache reads it
costs 9.9\%. It is cheaper than \texttt{perf record -e 'io\_uring:*'} on all
three (2.3\%, 7.0\%, and 13.3\%), while delivering reconstructed per-request
state rather than a raw event file that needs post-processing. The reason is
headroom. When the device is the bottleneck, tracepoint cost is absorbed
into CPU idle time instead of displacing I/O work.

Second, at CPU-bound saturation, in-kernel aggregation is not free, and we
report that plainly. Against \texttt{null\_blk} at 810k IOPS,
\tool-aggregate costs 31.9\% of throughput, or 583\,ns per request. That is
the price of an in-flight hash insert and delete plus a histogram update on
every submission and completion pair. A count-only \texttt{bpftrace} probe
costs 42\,ns (3.3\%) but answers no question beyond ``how many.''
\texttt{perf} costs 837\,ns (40.2\%) and defers all semantics to offline
processing, and \tool-trace costs 1{,}221\,ns (49.5\%). The naive claim that
in-kernel aggregation makes observation free at any load is false, and our
data says so. The defensible claim is that cost per unit of fidelity is the
right metric, and on that metric \tool-aggregate dominates the alternatives
that can actually answer latency-attribution questions.

Third, where the cost lands also differs. Whole-system CPU accounting shows
\tool-aggregate within run-to-run noise of baseline (7063 versus 7050
jiffies), because its work runs inline in the traced task and is already
counted in that task's time. By contrast, \texttt{perf} and \tool-trace
roughly double system CPU (13.8k and 15.1k jiffies), since each burns an
additional core on a userspace collector. On a machine with spare cores
this is invisible. On a saturated one, per-event collectors steal a core
from the workload's neighbors. \texttt{strace} is the anti-baseline at
44.4\,$\mu$s per request and 93 to 97\% throughput loss. The one exception
is SQPOLL, where it costs only 0.2\% because there are no syscalls to
intercept. Its overhead reaches zero exactly when its information content
does.

\subsection{Fidelity}
Across the grid, \tool{} reconstructed 100\% of completed requests in both
modes: in the heaviest cell, 36{,}838{,}423 submissions against
36{,}838{,}148 completions with 275 legitimately in flight at detach, zero
untracked completions, and zero ring-buffer drops as counted by trace mode's
drop counter (\texttt{trace\_rb\_drops}), which the report surfaces
directly. Trace mode sustained full
per-request capture at 420k IOPS (28.6M requests in one 68\,s cell). The
fidelity comparison with the alternatives is categorical rather than
numeric. The count-only probe reconstructs nothing. \texttt{perf} captures
raw events whose request-level correlation is deferred to offline tooling.
And \texttt{strace} cannot see the request flow at all.

\subsection{Tail perturbation}
\begin{figure}
\centering
\includegraphics[width=\columnwidth]{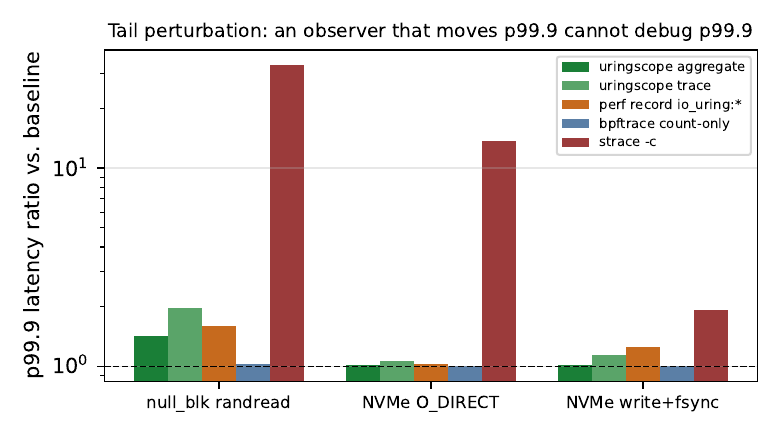}
\caption{p99.9 latency under observation, relative to baseline (log
scale). An observer that moves the tail cannot be used to debug the tail.}
\label{fig:tail}
\end{figure}
Figure~\ref{fig:tail} shows the tail. On device-bound NVMe workloads every
eBPF-based observer holds p99.9 within $1.06\times$ of baseline, and
\tool-aggregate stays within $1.02$ to $1.03\times$. At CPU-bound saturation
\tool-aggregate reaches $1.42\times$ and \texttt{perf} reaches $1.59\times$.
\texttt{strace} distorts p99.9 by $33\times$, which disqualifies it from tail
investigation even where its throughput cost might be tolerated.

\subsection{Portability, exercised live}
The portability mechanism was tested by events, not just by CI. The
measurement host runs Linux 6.17, released after this tool was written,
and 6.17 changed \texttt{io\_uring\_complete}'s prototype again. The scalar
\texttt{user\_data}, \texttt{res}, and \texttt{cflags} arguments collapsed
into a single \texttt{struct io\_uring\_cqe *}. \tool's probe detected the
unrecognized 3-argument form and selected it at load time, after one added
program variant of about 40 lines. The startup support matrix on this kernel
reports all nine lifecycle features active, with completion served by the
``v6.17 cqe-collapsed (3-arg)'' variant. The
same campaign also validated the 5.19 \texttt{io\_kiocb} field-move
flavor (poisoned on this kernel, as designed) and degrade-not-abort
behavior. Tracepoint format dumps from the measurement kernel are
archived as the churn table's primary sources.

\subsection{End-to-end boundary instrumentation}
\label{sec:eval-e2e}
The two userspace-side lifecycle segments (\S\ref{sec:lifecycle}) are
reached with best-effort \texttt{liburing} uprobes. On the measurement
host the tool located \texttt{liburing.so.2}, attached, and reported the
submit-side gap directly. For a workload submitting through the library, it
reported \emph{``1 io\_uring\_submit() call, avg 2.0 SQEs pending/call.''}
The reap-side gap, however, was not observable for our workloads, and we
report this honestly. \texttt{liburing}'s completion-peek path is frequently
inlined, emitting no library call for a uprobe to attach to, so reap-lag
timing is unavailable unless the application reaps through a non-inlined
entry point. This is the predicted asymmetry. Kernel-side CO-RE gives
one-binary portability, while uprobe attachment depends on the target
library's symbols and inlining. The end-to-end features are therefore
explicitly a best-effort tier that degrades to kernel-side-only with a clear
message rather than failing. The \texttt{[REAP-LAG]} rule is nonetheless
validated by injection (\S\ref{sec:eval-detect}), and what varies by target
is whether a uprobe site exists to feed it.

\subsection{Pathologies are kernel-version-dependent}
\label{sec:eval-pathology-drift}
We also had an unplanned finding. The classic ``buffered-read punt storm,''
the io\_uring pathology most documented by practitioners, does not reproduce
on 6.17. Across 14.27M cold-cache buffered reads, 0.0\% punted, because
modern kernels retry buffered reads via page-lock waiting rather than io-wq.
Buffered writes and fsync, by contrast, still punt 100\% (3.03M ops in our
fsync workload, attributed per-op as WRITE p50 262\,$\mu$s, FSYNC p50 1.0\,ms,
p99 4.2\,ms). The pathology catalogue itself drifts with the kernel. This is
an argument for a maintained measurement tool over folklore, and it is why
\tool's doctor reports evidence rather than assuming which pathologies a
kernel can exhibit. The SQPOLL cells make the complementary point. We saw
1.46M IOPS with zero \texttt{io\_uring\_enter()} syscalls observed and the
poller 0.0\% idle, and the doctor correctly stays silent on a healthy
saturated ring.

\subsection{Case study: naming the pathology}
The detection campaign supplies the case study directly. On the fsync
workload the doctor attributes 100\% of 3.03M requests to io-wq punts and
splits the latency by opcode (WRITE p50 262\,$\mu$s versus FSYNC p50
1.0\,ms, p99 4.2\,ms), turning ``writes are slow'' into ``fsync detours
through the worker pool and owns your tail.'' On the iodepth-1 workload the
doctor flags batching at 1.00 SQEs per \texttt{io\_uring\_enter()} across
28.6M calls. That is syscall-per-op, the very thing io\_uring exists to
avoid.

\subsection{Detection effectiveness}
\label{sec:eval-detect}
The injection harness scores every injectable doctor rule against printed
ground truth on the measurement kernel (Linux 6.17.0-1017). All 11
injection-validated rules detect their injected pathology with zero false
positives: punt storm, batching
failure, CQ overflow, error floods, dropped requests, SQPOLL stall, io-wq
fan-out, overlapping in-flight buffers, registered-buffer-lifetime violation,
\texttt{munmap}-of-in-flight-buffer, and reaping lag (Table~\ref{tab:detect}).
The false-positive guards hold. A no-batch workload does not trip the punt
rule, routine worker counts do not trip the fan-out rule, and a clean
workload reports no hazard. This includes the
silent-corruption case, where two overlapping \texttt{READ\_FIXED}s into one
registered buffer both return success (\texttt{res}=4096) while one writer's
data is destroyed. The doctor names both \texttt{user\_data} tokens and the
overlapping range so the developer can locate the two requests in their own
code.

\begin{table}
\centering
\small
\begin{tabular}{@{}lll@{}}
\toprule
injected pathology & doctor rule & result \\
\midrule
io-wq punt storm        & \texttt{[WARN]} punt ratio   & detect \\
syscall-per-op          & \texttt{[INFO]} batching     & detect \\
CQ overflow             & \texttt{[CRIT]} overflow     & detect \\
error flood             & \texttt{[WARN]} error rate   & detect \\
dropped requests        & \texttt{[LEAK]}              & detect \\
SQPOLL stall            & \texttt{[WARN]} sqpoll       & detect \\
io-wq fan-out           & \texttt{[WARN]} workers      & detect \\
overlapping in-flight   & \texttt{[HAZARD]}            & detect \\
reg-buffer lifetime     & \texttt{[HAZARD-BUFREG]}     & detect \\
\texttt{munmap} in-flight & \texttt{[HAZARD-UAF]}      & detect \\
reaping lag             & \texttt{[REAP-LAG]}          & detect \\
\midrule
\multicolumn{2}{l}{false-positive guards (clean workloads)} & 0 fired \\
\bottomrule
\end{tabular}
\caption{Detection effectiveness on Linux 6.17.0-1017: 11/11 injected
pathologies detected, 0 false positives. Scored by the injection harness
against machine-readable ground truth.}
\label{tab:detect}
\end{table}

\section{Related Work}

The closest prior tool is \texttt{uring-trace}~\cite{uring-trace}, which
also uses eBPF with Perfetto output. It is per-event by design and pinned to
kernels 6.1 through 6.7. Practitioner guides~\cite{cloudflare-iowq,
kernel-iouring-tracepoints} document manual tracepoint analysis, and the
strace-devel thread~\cite{strace-iouring-thread} records the gap without
closing it.

On portability, DepSurf~\cite{depsurf} measures the eBPF dependency-surface
instability that \S\ref{sec:churn} engineers around, and CO-RE~%
\cite{libbpf-core} provides the field-offset half of the solution.
RingGuard~\cite{ringguard} combines io\_uring with eBPF, but for security
policy enforcement rather than observability or correctness. More generally,
bcc/libbpf-tools and \texttt{bpftrace}~\cite{bpftrace} provide the substrate
that \tool{} builds on. \tool{} differs in being io\_uring-specific,
semantics-aware, and packaged as a single binary, and it uses
Perfetto~\cite{perfetto} as its trace-mode output target.

\section{Limitations and Future Work}
\label{sec:limitations}

\tool{} has real limits. The two userspace-side lifecycle segments
(SQE-prep to submit, and CQE-ready to reap) are observed with best-effort
\texttt{liburing} uprobes, and their visibility depends on how the
application reaps (\S\ref{sec:eval-e2e}). The submit side is observable when
submission goes through the library. The reap side is observable only when
the application reaps through a non-inlined library call or a waiting
\texttt{io\_uring\_enter}. Applications that inline the completion-peek path
or read the completion ring directly expose no instrumentation point, so for
them completion-reaping lag is unmeasured. The 5.15 legacy tier is
counters-only, without per-request punt attribution.
io\_uring grows opcodes and semantics every cycle (zero-copy receive, futex
operations, bundles), so the lifecycle model requires ongoing maintenance.
The overlapping-in-flight check is bounded to the last $K$ requests per ring.
The seccomp and container reality (\S\ref{sec:background}) limits the
deployment surface by construction, and the measurement host being
virtualized (\S\ref{sec:setup}) leaves a clean bare-metal re-run as future
work.

\section{Conclusion}

The packaged-io\_uring-observability slot stood empty not for lack of data,
which the kernel has exposed via tracepoints for years, but because the
surface was hostile: the request flow lives in shared memory, and the
tracepoints are not ABI. A precise lifecycle model and a portability
discipline of BTF-probed variants, CO-RE flavors, and struct-centric reads
close the gap. The result is a single static binary that observes any
language's io\_uring traffic on stock kernels from 5.15 through 6.17, costs
0.7--9.9\% on device-bound workloads, reconstructs every request, names
pathologies with evidence, and adapted to a kernel released after it was
written. \tool{} is open source.

\bibliographystyle{ACM-Reference-Format}
\bibliography{refs}

\end{document}